\documentclass[twocolumn]{aastex63}

\usepackage[english]{babel}
\usepackage{amsmath}
\usepackage{graphicx}
\usepackage{wrapfig}
\usepackage[colorinlistoftodos]{todonotes}
\usepackage{natbib}

\newcommand{\hcop}{HCO$^+$}

\newcommand{\hcisoop}{H$^{13}$CO$^+$}
\newcommand{\nthp}{N$_2$H$^+$}
\newcommand{\ang}{\mbox{\normalfont\AA}}

\begin{document}

\title{Evidence for a Cosmic Ray Gradient in the IM Lup Protoplanetary Disk}

\author{Richard A. Seifert}
\correspondingauthor{Richard A. Seifert}
\email{ras8qnr@virginia.edu}
\affiliation{Department of Astronomy, 
University of Virginia, 
Charlottesville, VA 22904, USA}
\author[0000-0003-2076-8001]{L. Ilsedore Cleeves}
\affiliation{Department of Astronomy, 
University of Virginia, 
Charlottesville, VA 22904, USA}

\author[0000-0002-8167-1767]{Fred C. Adams}
\affiliation{Department of Physics, 
University of Michigan, 
Ann Arbor, MI 48109, USA}
\affiliation{Department of Astronomy, 
University of Michigan, 
Ann Arbor, MI 48109, USA}

\author{Zhi-Yun Li}
\affiliation{Department of Astronomy, 
University of Virginia, 
Charlottesville, VA 22904, USA}

\date{\today}

\begin{abstract} 
Protoplanetary disk evolution is strongly impacted by ionization from the central star and local environment, which collectively have been shown to drive chemical complexity and are expected to impact the transport of disk material. Nonetheless, ionization remains a poorly constrained input to many detailed modeling efforts. We use new and archival ALMA observations of \nthp{} 3--2 and \hcisoop{} 3--2 to derive the first observationally-motivated ionization model for the IM Lup protoplanetary disk. Incorporating ionization from multiple internal and external sources, we model \nthp{} and \hcisoop{} abundances under varying ionization environments, and compare these directly to the imaged ALMA observations by performing non-LTE radiative transfer, visibility sampling, and imaging. We find that the observations are best reproduced using a radially increasing cosmic ray (CR) gradient, with low CR ionization in the inner disk, high CR ionization in the outer disk, and a transition at $\sim 80 - 100$ au. This location is approximately coincident with the edge of spiral structure identified in millimeter emission. We also find that IM Lup shows evidence for enhanced UV-driven formation of \hcop{}, which we attribute to the disk's high flaring angle. In summary, IM Lup represents the first protoplanetary disk with observational evidence for a CR gradient, which may have important implications for IM Lup's on-going evolution, especially given the disk's young age and large size.
\end{abstract}


\section{Introduction}
Ionization plays a crucial role in the evolution of protoplanetary disks, directly impacting both their chemical and physical evolution. Ions are a key factor in driving up chemical complexity, particularly through the liberation of atomic hydrogen from molecular H$_2$ which ultimately leads to the formation of water \citep{vandishoeck2013} and organics \citep{cleeves16}. The presence of ions also plays a role in disk physical evolution by coupling the gas to magnetic fields. This coupling allows for the redistribution of angular momentum and the emergence of turbulence driven by the magneto-rotational instability (MRI; \citealt{balbus91}). Turbulence, in turn, influences the formation of planetesimals by setting the distribution of mass in disks and by governing the growth of dust grains \citep[e.g.,][]{dullemond2007,testi2014}.

\begin{figure*}[t]
   \begin{center}
    \includegraphics[width=1.0\textwidth]{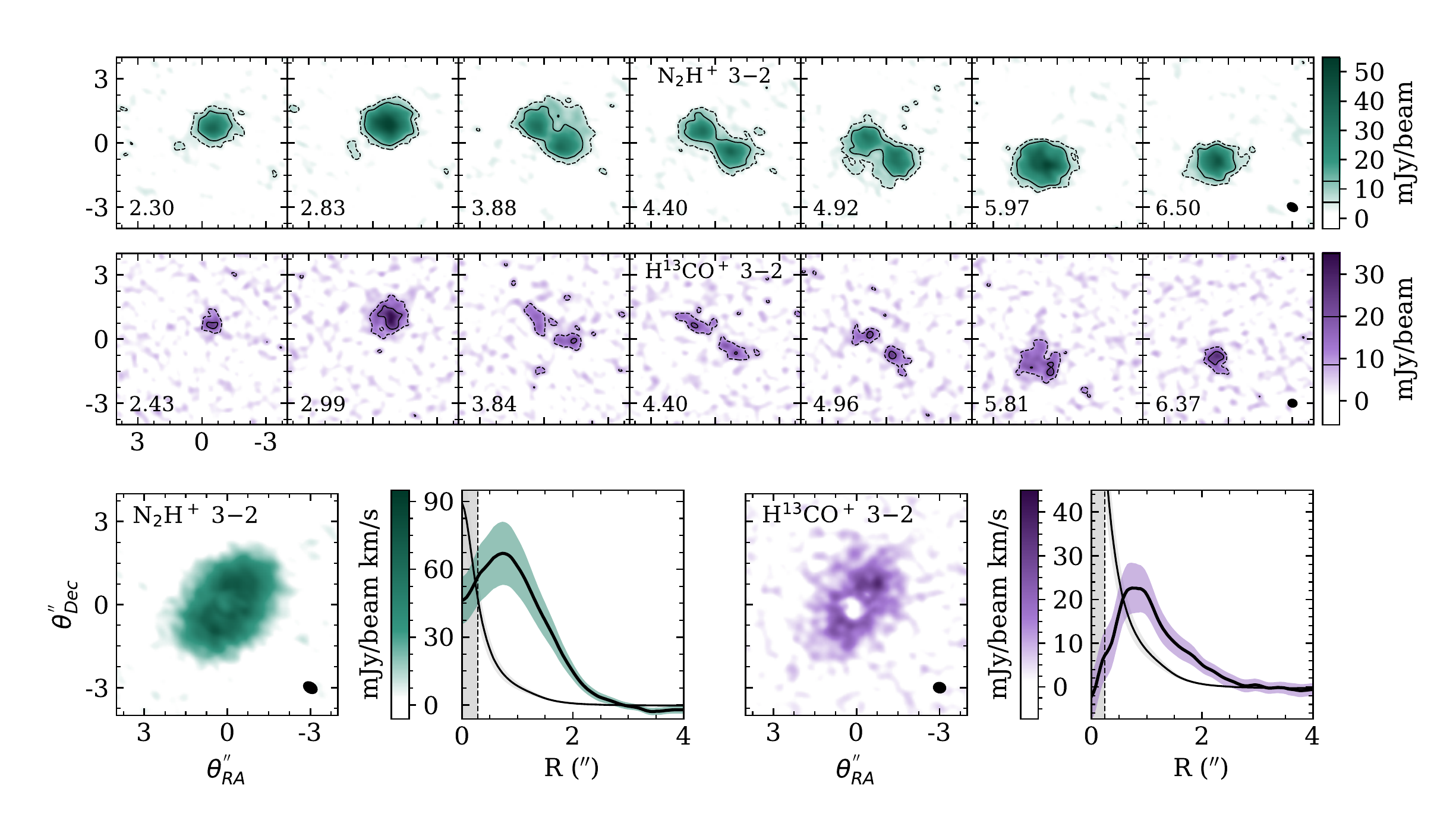}
    \vspace{-8mm}
    \label{fig:channels}
    \caption{Channel maps for \nthp{} 3--2 (top) and \hcisoop{} 3--2 (middle), with moment-0 maps and radial intensity profiles for both (bottom). \textbf{Channel Maps:} Dotted and solid contours represent 3$\sigma$ and 7$\sigma$ respectively. The beam is indicated in the bottom right corner, and the velocity relative to the local standard of rest is indicated in km s$^{-1}$ in the bottom left corner for each channel. \textbf{Radial Profiles:} The central beam semi-major axis is indicated with gray shading. In addition to line radial profiles, 1 mm continuum radial intensity is also overlaid in both panels. See Section \ref{sec:singlecr} for a description of radial intensity and uncertainty calculations.}
    \vspace{-5mm}
    \end{center}
    \end{figure*}
    
\begin{table*}[]\label{tab:obs}
    \begin{center}
    \caption{Line and Continuum Observations}
    \begin{tabular*}{\textwidth}{l@{\extracolsep{\fill}}ccccc}
         \hline \hline
       Transition  & Rest Frequency & Beam (Position Angle) & Channel Width & RMS per beam & Disk-Integrated Flux$^a$ \\
       & (GHz) & & (km/s) & (mJy / beam) &  (Jy km s$^{-1}$) \\ \hline
       \nthp{} 3--2 & 279.512 & 0$^{''}$\hspace{-1.5mm}.57$\times$0$^{''}$\hspace{-1.5mm}.42 (57.8$^{\circ}$) & 0.524 & 1.81 & 1.6 $\pm$ 0.18 \\
       \hcisoop{} 3--2 & 260.255 & 0$^{''}$\hspace{-1.5mm}.49$\times$0$^{''}$\hspace{-1.5mm}.42 (77.7$^{\circ}$) &  0.282 & 2.85 & 0.64 $\pm$ 0.067 \\ \hline
       Continuum & 1 mm & 0$^{''}$\hspace{-1.5mm}.57$\times$0$^{''}$\hspace{-1.5mm}.42 (57.8$^{\circ}$) & 14.6 kHz & 0.228 & 0.33 $\pm$ 0.053 Jy\\ \hline
    \end{tabular*}
    \end{center}
    \textbf{Note.}\newline$^a$ Computed within a Keplerian mask centered on the disk. Errors estimated by applying the mask to line-free channels, with additional 10\% flux uncertainty. Continuum flux computed in an 8$''$ box.
    \label{tab:my_label}
\end{table*}

Ionization in the disk can come from many sources: the star provides UV and X-ray photons and may produce locally accelerated cosmic rays (CRs) in shocks along magnetically launched jets or at accretion cites on the stellar surface \citep{padovani2020}; the disk provides $\gamma$-rays and $\beta$-particles emitted by short-lived radionuclides (RNs) such as $^{26}$Al and $^{60}$Fe \citep{umebayashi81,finocchi1997,cleeves13b}; and the external environment provides additional interstellar UV photons \citep[e.g.,][]{fatadams,adams2010} and cosmic rays \citep{cleeves13a}.

Each ionizing agent interacts with the disk in a different way and leaves distinctive chemical signatures within the disk \citep[e.g.,][]{cleeves14a}. UV photons ionize the surface layers of the disk but are limited by small grain dust opacity. X-ray photons can ionize deeper regions of the disks (depending on the hardness of the stellar X-ray spectrum). Radioactive materials can dominate the ionization in regions near the disk midplane. And cosmic rays --- which can have sufficient energies to penetrate down to the midplane --- ionize all layers of the disk up to gas column densities of $\sim 100$ g~cm$^{-2}$ \citep{umebayashi81,cleeves13a,padovani2018}.

This simple picture of ionization in disks is complicated by a number of uncertainties regarding the amounts and distributions of each source of ionization, including uncertainties in the hardness of the stellar X-ray spectrum and in the abundance and distribution of RNs across the midplane, as well as the unknown degree of CR exclusion near the star due to winds and/or magnetic deflection \citep{cleeves13a}. However, with knowledge of the density and temperature structure of the disk, and by utilizing the strong relationship between ionization and disk chemistry, the level of ionization in different regions of the disk can be inferred from the abundances and distributions of a few key ionization-tracing molecules. Such efforts are analogous to the use of molecular ions as probes of ionization in star forming regions \citep[e.g.,][]{caselli2002}. 

In this work, we use the historically successful ionization tracers \hcisoop{} and \nthp{} \citep{caselli2002,cleeves14a,cleeves15,quenard2018} to place constraints on the radially resolved ionization structure in the IM Lup protoplanetary disk. IM Lup is an M0 star in the Lupus 2 cloud, which sits at roughly 161 pc away. The star has a luminosity of L$_*$ = 0.9 L$_\odot$ \citep{hughes94} and a mass of M$_*$ = 1 M$_\odot$ \citep{panic09,cleeves16}. Its age is estimated to be $0.5-1$ Myr \citep{mawet12}. Based on both {\em Hubble Space Telescope} scattered light observations and SMA resolved millimeter continuum observations, the disk around IM Lup is estimated to have a total mass of $\sim$0.1 M$_\odot$ and a large size, at least 600~au in radius \citep{pinte08,panic09,cleeves16,avenhaus,pinte18}. Therefore IM Lup is an excellent test bed to not only look at ionization fractions, but also to better understand its spatial distribution to disentangle the importance of the different ionizing agents.

The paper is laid out as follows. In Section~\ref{sec:obs} we describe the observations used to constrain the radial ionization profile of IM Lup. In Section~\ref{sec:model} we describe the underlying disk physical structure, ionization models considered, the chemical modelling procedure, and the generation of synthetic observations. Section~\ref{sec:results} describes findings from our analysis of dominant chemical pathways for the observed ions, and presents an overview of requisite ionization conditions to explain the observed emission. Finally, in Section~\ref{sec:disc} we discuss the implications of our findings and in Section~\ref{sec:conclusions} we summarize our conclusions.

\section{ALMA Observations}
\label{sec:obs}


The observations presented in this work were carried out with ALMA as part of Project Code 2013.1.00694.S (PI: Cleeves), augmented by archival data of H$^{13}$CO$^+$ 3--2 from the program 2013.1.00226 (PI: {\"Oberg}). The H$^{13}$CO$^+$ 3--2 calibration was previously reported in \citet{cleeves2017} and \citet{oberg15}, and for details we point the reader to those papers. Since \citet{cleeves2017} find that three epochs of H$^{13}$CO$^+$ 3--2 observations toward this source are variable, we only use the two lower and consistent flux epochs of 2014 July 17th and 2015 January 29th \citep[see ][for further details]{cleeves2017}.

The Band 7 N$_2$H$^+$ 3--2 data presented here has not been previously reported. The extended N$_2$H$^+$ 3--2 observations were carried out on 2014 July 8 with 33 antennas spanning 20m -- 650m baselines. The data were initially calibrated via the CASA Pipeline. Titan was used as the amplitude calibrator, J1427-4206 as the bandpass calibrator, and J1534-3526	as the phase calibrator. The compact N$_2$H$^+$ 3--2 observations were carried out on 2014 December 24 with 40 antennas spanning 15m -- 349m baselines. These observations were also calibrated by the CASA Pipeline with Titan as the amplitude calibrator, J1427-4206 as the bandpass calibrator, and J1610-3958 as the phase calibrator. 

For both data sets, we applied additional phase self-calibration combining polarizations and across scans. We also set a minimum SNR for solutions of 3 and a minimum baseline per antenna of 6. We self-calibrated the extended and compact data separately before imaging them together. For the extended data, we applied two rounds of phase self calibration on the continuum with a solution interval of infinity and then 30s, yielding a factor of 2.8$\times$ improvement on the RMS of the continuum data. For the compact data we applied only one round of phase self calibration with a solution interval of infinity and obtained a 6$\times$ improvement on the RMS noise of the continuum. Solutions were applied to the lines separately to each spectral window.

Continuum subtraction and imaging was carried out using the \texttt{uvcontsub} and \texttt{tclean} tasks in CASA 5.6.1. We adopted a briggs weighting with a  \texttt{robust} parameter of 0.5 to image both lines, resulting in beams of 
(0$^{''}$\hspace{-1.5mm}.57$\times$0$^{''}$\hspace{-1.5mm}.42, 57.8$^{\circ}$) 
and 
(0$^{''}$\hspace{-1.5mm}.49$\times$0$^{''}$\hspace{-1.5mm}.42, 77.7$^{\circ}$), for \nthp{} 3--2 and \hcisoop{} 3--2, respectively. A subset of channel maps as well as moment-0 maps and radial intensity profiles for both lines are presented in Figure \ref{fig:channels}, and additional information on our observations is shown in Table \ref{tab:obs}. From these images, we see that both ions have inner deficits roughly 1$^{''}$ in radius. This is about twice the size of a single beam and is larger than the region of high dust optical depth found in \citet{huang18}, indicating that inner deficits in emission are not fully explained by optical depth effects from dust continuum.

\section{Modeling IM Lup}\label{sec:model}

In the following sections, we aim to explain the observed \nthp{} and \hcisoop{} emission distribution using astrochemical models applied to a fixed physical density and temperature structure (Section~\ref{sec:phys}) with varying underlying ionization assumptions (Section~\ref{sec:ionmod}). For each ionization model, disk chemical models are computed (Section~\ref{sec:chemmod}), and the output is compared the data via synthetic observations (Section~\ref{sec:synth}).

\subsection{Physical Model}\label{sec:phys}
The IM Lup physical structure is taken from \citet{cleeves16} and is described in detail in Sections 3.1 and 3.2 therein. The model was developed using ALMA CO and continuum visibilities and a dust spectral energy distribution. The gas density follows the standard self-similarity solution of \citet{lyndenbell74} presented in \citet{andrews11}. The dust disk is modeled as a combination of two populations: a small grain population (0.005 \micron{} - 1 \micron{}) cospatial with gas extending out to a radius of 970 au, and a large grain population (0.005 \micron{} - 1 mm) spatially decoupled from the gas and sharply truncated at a radius of 313 au. Both grain populations follow an MRN size distribution \citep{mathis77}.

The density structure of the gas and dust and the thermal structure of the dust were found using the TORUS code \citep{harries00,harries04,kurosawa09,pinte09}, assuming passive disk heating by a central star of effective temperature $T_{\rm eff}$ = 3900 K and radius R$_*$ = 2.5 R$_{\odot}$ \citep{pinte08}. Gas temperatures were estimated using a fitting function from \citet{bruderer13}, which parameterizes gas-dust thermal decoupling given local UV flux and gas density.

\begin{figure*}[t]
  \begin{center}
    \includegraphics[width=1.07\textwidth]{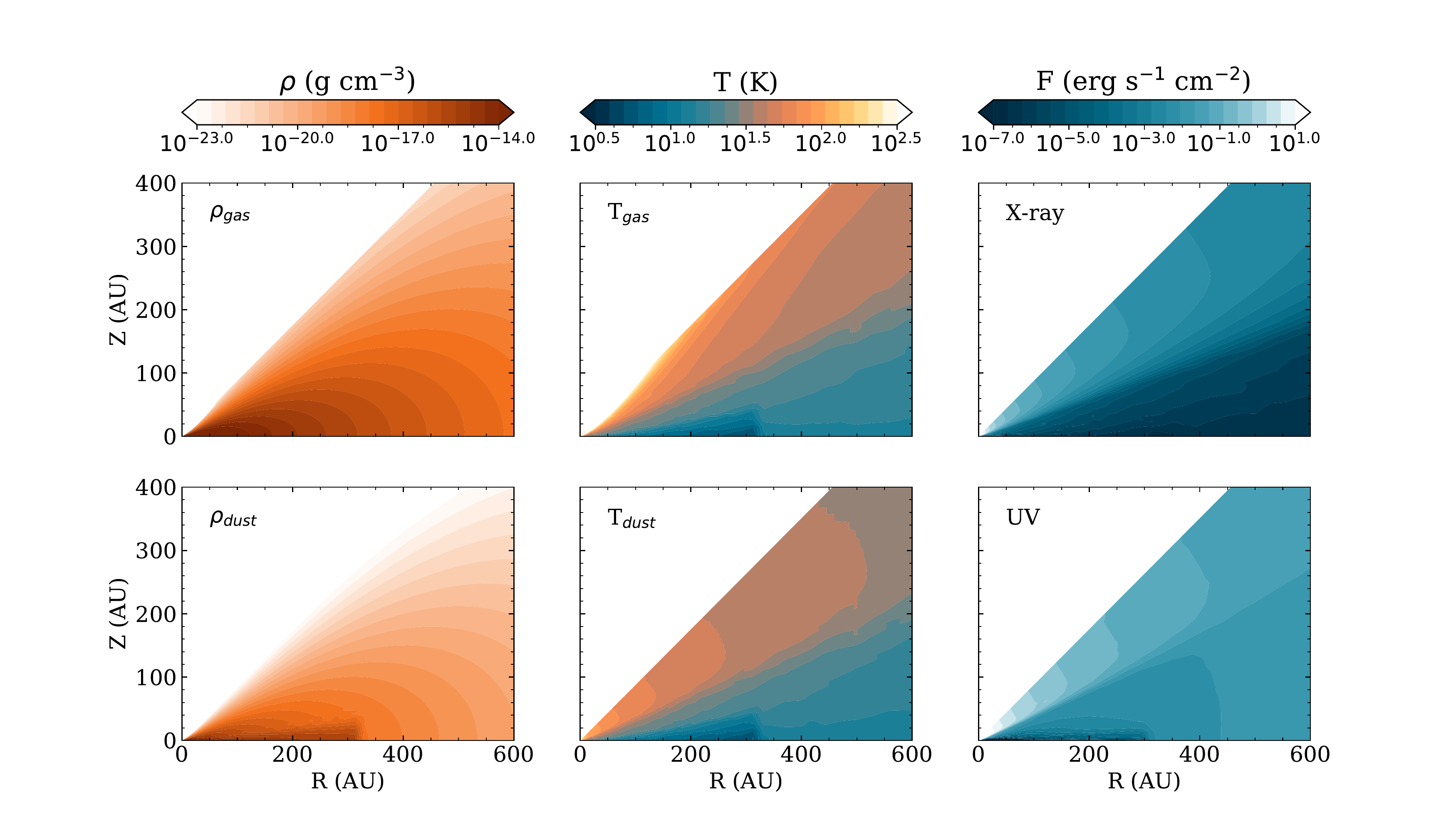}
    \caption{Physical model of IM Lup as adopted from \citet{cleeves16}. Note, $\rho_{\rm dust}$ includes both dust density in small grains and large grains, hence the visible concentration near the midplane inside of 300 au. \label{fig:phys}}
\end{center}
\end{figure*}

\subsection{Ionization Model}\label{sec:ionmod}

Our model incorporates ionization from UV and X-ray photons and CRs. We utilize existing constraints on UV and X-ray emission, and use the Monte Carlo radiation transfer code from \citet{bethell11} to obtain two dimensional UV and X-ray fluxes for our disk model. We consider a variety of different scenarios for the CRs, and utilize the specific CR models of \citep{cleeves13a} for this work.

Constraints on FUV (912-2000\ang{}) come from the \textit{Neil Gehrels Swift Observatory} using the UVM2 filter. As described in \citet{cleeves16}, the detailed {\em HST} FUV spectum of TW Hya \citep{herczeg02,herczeg04} is adopted as a template, which we then re-scale according to the \textit{Swift} UVM2 IM Lup and TW Hya measurements to obtain a model FUV spectrum for IM Lup. The X-ray spectrum of IM Lup has been updated from the models presented in \citet{cleeves16} to incorporate IM Lup's \textit{Swift} XRT observed X-ray spectrum as described in \citet{cleeves2017} (see their Figure 2).

For the CR component, we use six different model templates based on the work of \citet{cleeves13a}. These models incorporate exclusion of low energy CRs to varying degrees, by approximating the impact of magnetized wind-driven suppression (from the star or disk), as well as energy decay with depth. We consider two ISM analogue models: The first one from \citealt{moskalenko02} (M02) approximates the diffuse ISM, whereas the second from \citealt{webber98} (W98) approximates the dense molecular ISM. We also test four circumstellar analogue models: The Solar System Minimum (SSM) and Solar System Maximum (SSX) models match current CR rates on Earth at 1 au, and thus bracket the expected modulation behavior for our solar system. The T Tauri Minimum (TTM) and T Tauri Maximum (TTX) models extrapolate the solar wind outflow rates to those expected for T Tauri stars and thus produce enhanced modulation of CRs. A summary of the model properties is shown in Table \ref{tab:CRmods} (for additional details on these CR templates, see \citealt{cleeves13a}).

\begin{table}[t]
    \caption{Model incident CR ionization rates assumed integrated over CR energy. For dependence with column density see \citet{cleeves13a}.}
    \begin{tabular}{lcr}
         \hline\hline
         Model & ID & $\zeta_{\rm CR}$ (s$^{-1}$) \\ 
         \hline
         \citealt{moskalenko02} & M02 & 6.8 $\times$ 10$^{-16}$ \\
         \citealt{webber98} & W98 & 2.0 $\times$ 10$^{-17}$ \\
         Solar System Min & SSM & 1.1 $\times$ 10$^{-18}$ \\
         Solar System Max & SSX & 1.6 $\times$ 10$^{-19}$ \\
         T Tauri Min & TTM & 7.0 $\times$ 10$^{-21}$ \\
         T Tauri Max & TTX & 1.0 $\times$ 10$^{-21}$ \\
         \hline
    \end{tabular}
 \label{tab:CRmods}
\end{table}

\subsection{Chemical Model}\label{sec:chemmod}

Abundances of \nthp{} and \hcisoop{} are computed using the full 2D time-dependent chemical code presented in \citet{fogel11} and modified by \cite{cleeves14a}. The physical structure and radiation models described above and shown in Figure \ref{fig:phys} are used to calculate abundances as a function of position and time for 643 chemical species subject to 5976 reactions. The reaction network we use is based on the Ohio State University gas-phase network presented in \citet{smith04} with the addition of grain surface reactions and self-shielding. Only two-body reactions are considered, including ion-neutral, neutral-neutral, ion dissociative recombination, photon-driven chemistry, freeze-out, thermal and non-thermal desorption, grain surface chemistry, and self-shielding of CO, N$_2$, and H$_2$. The network adopted in this work does not include deuterium or other isotopes, thus we use a constant $^{12}$C/$^{13}$C factor of 60 \citep{langer1993} to obtain \hcisoop{} abundances. Given that isotope-selective self shielding has a stronger impact on the isotopic precursor C$^{18}$O compared to $^{13}$CO \citep[e.g.,][]{miotello2014}, we do not expect this simplification to have a major impact on our results.

\begin{figure*}[t]
 \begin{center}
\includegraphics[width=1.0\textwidth]{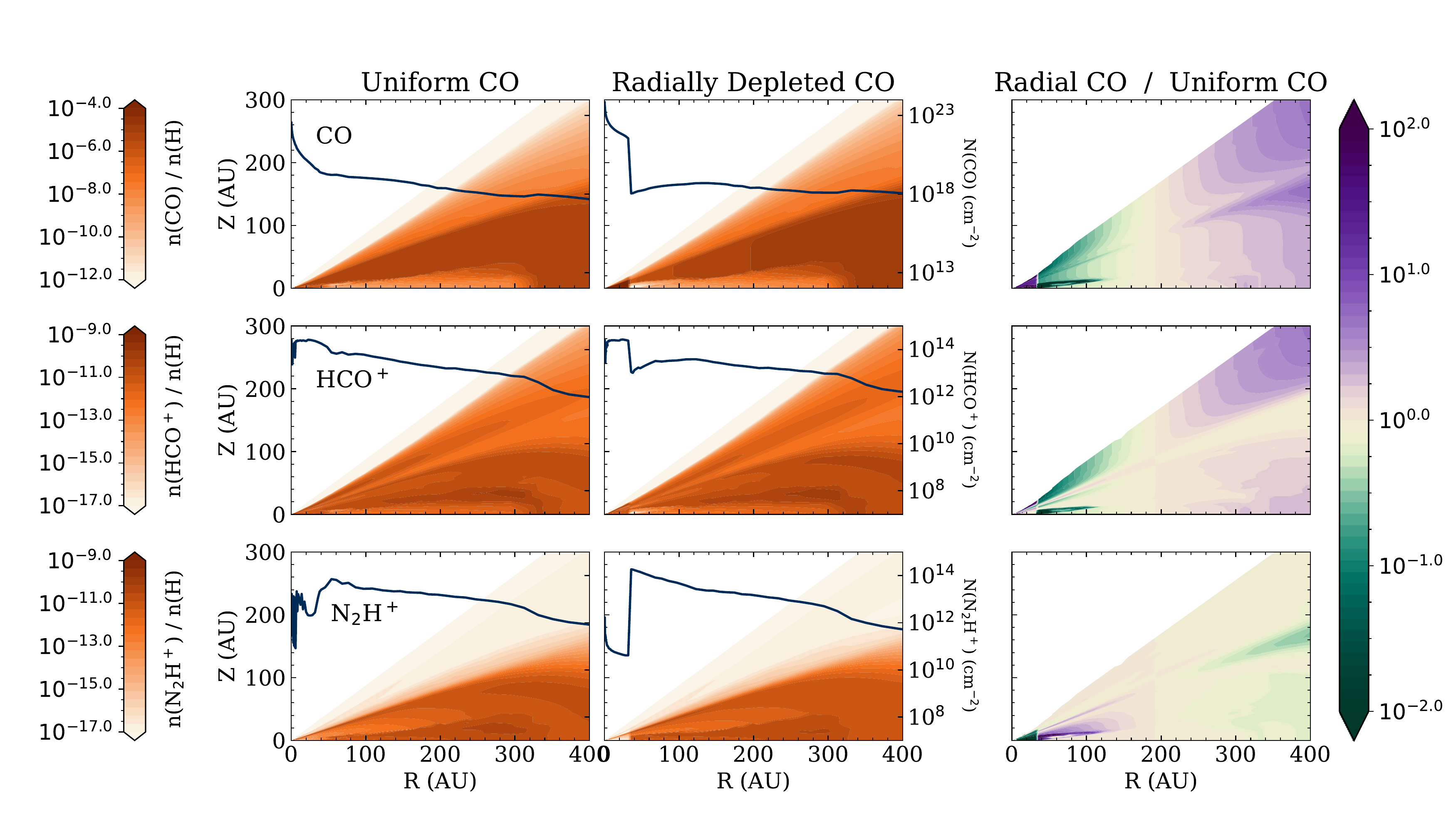}
\caption{Model abundances after 0.5 Myr for CO (top), \hcop{} (middle), and \nthp{} (bottom) assuming uniform (left) and radially depleted (center) initial CO distributions, with vertical column density indicated by dark blue curves. The ratio between uniform and radially depleted CO model abundances is shown in the right column for each of the three species. A radius of 175 au roughly corresponds to where both models have matching initial CO abundances.\label{fig:abundance}}
\end{center}
\end{figure*}

Initial abundances of the chemical species are chosen to approximate molecular cloud conditions, with most species initialized in the gas-phase except water, which is initialized on grains. Values we adopt are shown in Table \ref{tab:init}, and are based on those used by \citet{fogel11} with updated sulfur-bearing molecule abundances from \citet{cleeves14a} and IM Lup-specific updated H$_2$O(gr) and CO from \citet{cleeves18}, which found disk-averaged depletion factors of 50$\times$ and 20$\times$ for the two species, respectively. 

\begin{table}[t]
  \begin{center}
    \caption{Initial abundances for disk chemical modeling. All abundances are relative to total H density. Species are initialized in the gas phase except for grains and water ice. Initial electron density is computed based on the input ion abundances to ensure the simulation is charge neutral.}
    \begin{tabular}{lclc}
         \hline \hline
         Molecule & Abundance & Molecule & Abundance \\
         \hline
         H$_2$ & 5.00 $\times$ 10$^{-1}$ & He & 1.40 $\times$ 10$^{-1}$ \\
         CS & 4.00 $\times$ 10$^{-9}$ & CO & 7.00 $\times$ 10$^{-6}$ \\
         HCO$^+$ & 9.00 $\times$ 10$^{-9}$ & SO & 5.00 $\times$ 10$^{-9}$ \\
         N$_2$ & 3.75 $\times$ 10$^{-5}$ & H$_3^+$ & 1.00 $\times$ 10$^{-8}$ \\
         C$_2$H & 8.00 $\times$ 10$^{-9}$ & H$_2$O(gr) & 1.60 $\times$ 10$^{-6}$ \\
         Si$^+$ & 1.00 $\times$ 10$^{-9}$ & Mg$^+$ & 1.00 $\times$ 10$^{-9}$ \\
         Fe$^+$ & 1.00 $\times$ 10$^{-9}$ & Grains & 6.00 $\times$ 10$^{-12}$ \\
         \hline
    \end{tabular}
    \label{tab:init}
    \end{center}
\end{table}

We also explore models with radially dependent CO depletion. \citet{zhang19} find a radial gradient in the abundance of gas-phase CO in IM Lup with increasing CO depletion towards the disk center, which they attribute to migration of CO on dust grains as described in \citet{krijt18}. To incorporate this effect into our model, we approximate the radial CO depletion factor presented in \citet{zhang19} with a power law beyond the CO snowline (we use 32 au), and allow CO to return to ISM abundance interior to this. We then normalize the disk-wide CO depletion factor to retain the disk-integrated depletion factor of 20$\times$ from \citet{cleeves16}. We adopt a CO binding energy of 855~K \citep{oberg05}, and define the CO snowline as the radius at which half of CO is frozen on grains, which occurs at 32 au and a midplane temperature of 19~K in our models. A comparison of uniform CO and radially depleted CO models is shown in Figure \ref{fig:abundance}.

\subsection{Synthetic Observation Pipeline}
\label{sec:synth}

Chemical models are computed to an age of 0.5~Myr, an approximate age for IM Lup \citep{mawet12}. From the disk model abundances, we calculate the emergent flux from \nthp{} 3--2 and \hcisoop{} 3--2 using the LIME non-LTE radiation transfer code \citep{brinch10}. Collisional rates are taken from the LAMDA database \citep{schoier2005} for \nthp{} and \hcisoop{} \citep{Botschwina1993,flower1999}. We do not include hyperfine splitting and perform all simulations in non-LTE to account the possibility of emission arising in regions of low H$_2$ density.

All LIME simulations assume a distance of 161 pc \citep{gaia} and a disk inclination of 48$^\circ$ \citep{cleeves16}. A revised distance of 158 $\pm$ 3 pc was found in Gaia DR2, but as this distance is consistent with the DR1 distance, we choose to use the DR1 measurement so that our results can be directly compared with those from \citet{cleeves16}. Input gas velocities include Keplerian motion around the star, an isotropic turbulent velocity component of 100~m~s$^{-1}$, and thermal broadening. LIME produces sky brightness maps for discrete frequencies, so we spectrally over-sample by a factor of 40$\times$ and average down to the velocity resolution of the data in order to mimic the effects of channel smearing. We match the spectral resolution in the simulations to that of the observations: 0.524~km~s$^{-1}$ and 0.282~km~s$^{-1}$ for \nthp{}~3--2 and \hcisoop{}~3--2 respectively. We perform visibility sampling on the output LIME cubes using the code \texttt{vis\_sample} \citep{loomis18}. The original ALMA visibilities are provided as an input, allowing \texttt{vis\_sample} to construct a measurement set of the LIME output sky brightness maps with identical uv coverage as the data.

IM Lup exhibits an opaque central region attributed to optically thick dust within $\sim$20~au \citep{huang18}, unresolved in the present observations. For all ALMA gas observations of this source, the molecular emission is either depressed or absent in the region where the continuum is bright \citep[e.g.,][]{cleeves16}. It remains unclear how the dust and gas emission are interacting; whether optically thick dust is physically blocking gas emission, or continuum is being artificially subtracted from optically thick line emission, or both. Therefore this region poses a problem for directly assessing goodness-of-fit for our models in the visibility plane, since we are unable to selectively mask this problematic region. Instead, we have chosen to compare the goodness of fit between our models and data in the image plane, where the disk center can be explicitly ignored. To that end, we only compare the data and models outside of the central beam, beyond which the continuum in our models is optically thin everywhere (R~$\gtrsim$~40~au).

Synthetic measurement sets are continuum-subtracted, cleaned, and imaged in the same manner as described in Section \ref{sec:obs}. We found that, due to the faintness of \hcisoop{} emission in our models, cleaning of the noiseless model visibilities was highly susceptible to negative bowling, making models appear systematically under-bright compared to the data. To mitigate this effect, we artificially inject gaussian noise into our models corresponding to $\sim$10\% of the noise level of our data, which we found sufficient to prevent the negative bowling artifact during subsequent cleaning of the model visibilities.

\section{Results}\label{sec:results}

\subsection{\nthp{} and \hcisoop{} chemistry in IM Lup}\label{sec:chempath}

In the standard picture, \hcop{} and \nthp{} trace ionization in distinct vertical layers in the disk. This trend arises because both molecular ions are formed by H$_3^+$, and \nthp{} is efficiently converted to \hcop{} in the presence of CO. This behavior canonically leads to the emergence of four distinct regions in the disk: the ionized surface layer where CO and N$_2$ are photodissociated, a warm upper layer where CO is gaseous and \hcop{} probes ionization, a cool lower layer where CO is frozen out and \nthp\hspace{0cm} probes ionization, and the cold midplane where both CO and N$_2$ are frozen out \citep[e.g.,][]{qi19}.

\begin{figure}[t]
 \begin{center}
\includegraphics[width=.49\textwidth]{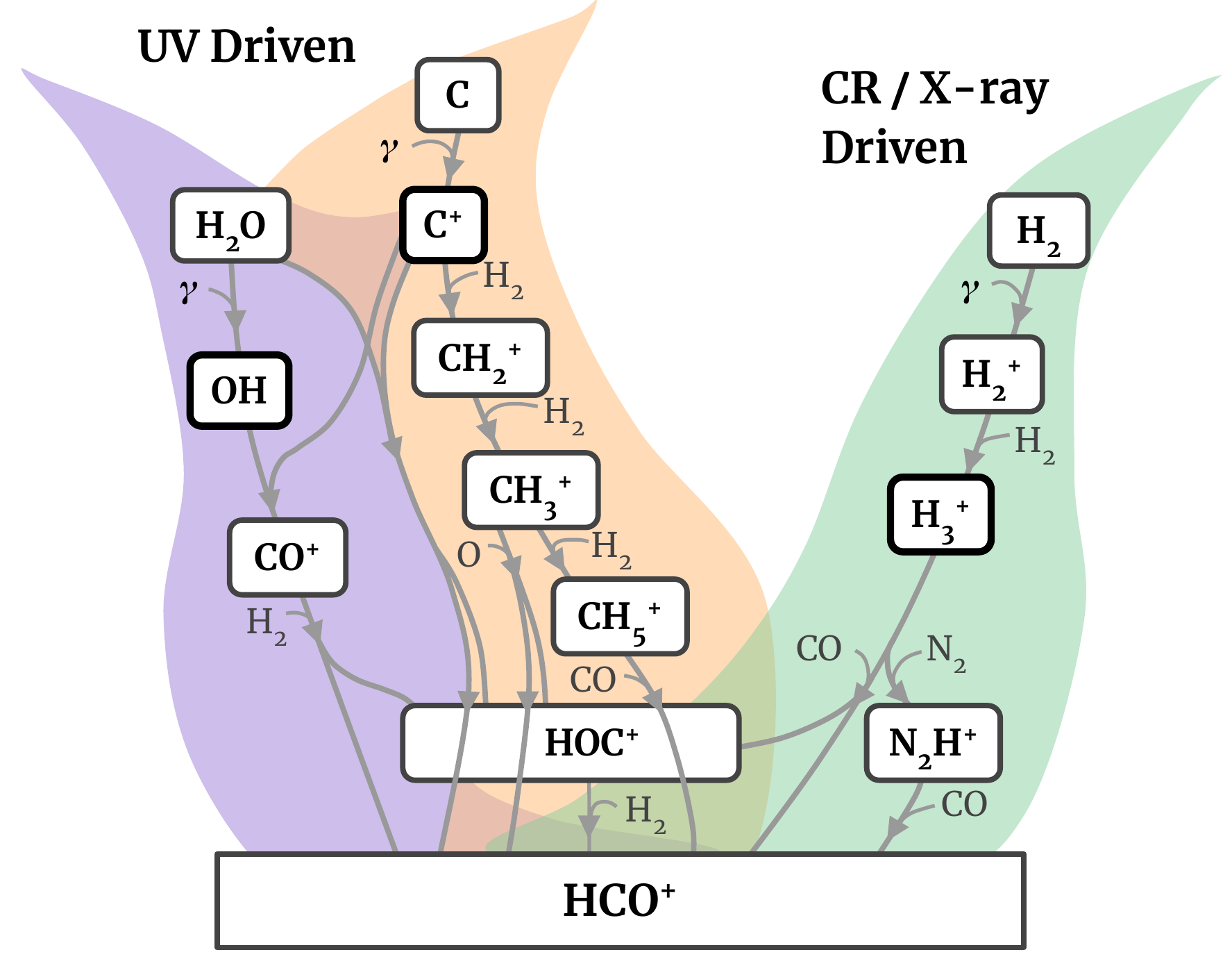}
\caption{Summary of the three distinct yet inter-related pathways dominating \hcop{} production in our chemical models of IM Lup. The contribution of each pathway as a function of location in the disk is shown in Figure \ref{fig:path_profiles}.}
\label{fig:path_diagram}
\end{center}
\end{figure}

\begin{figure}[t]
 \begin{center}
 
\includegraphics[width=.45\textwidth]{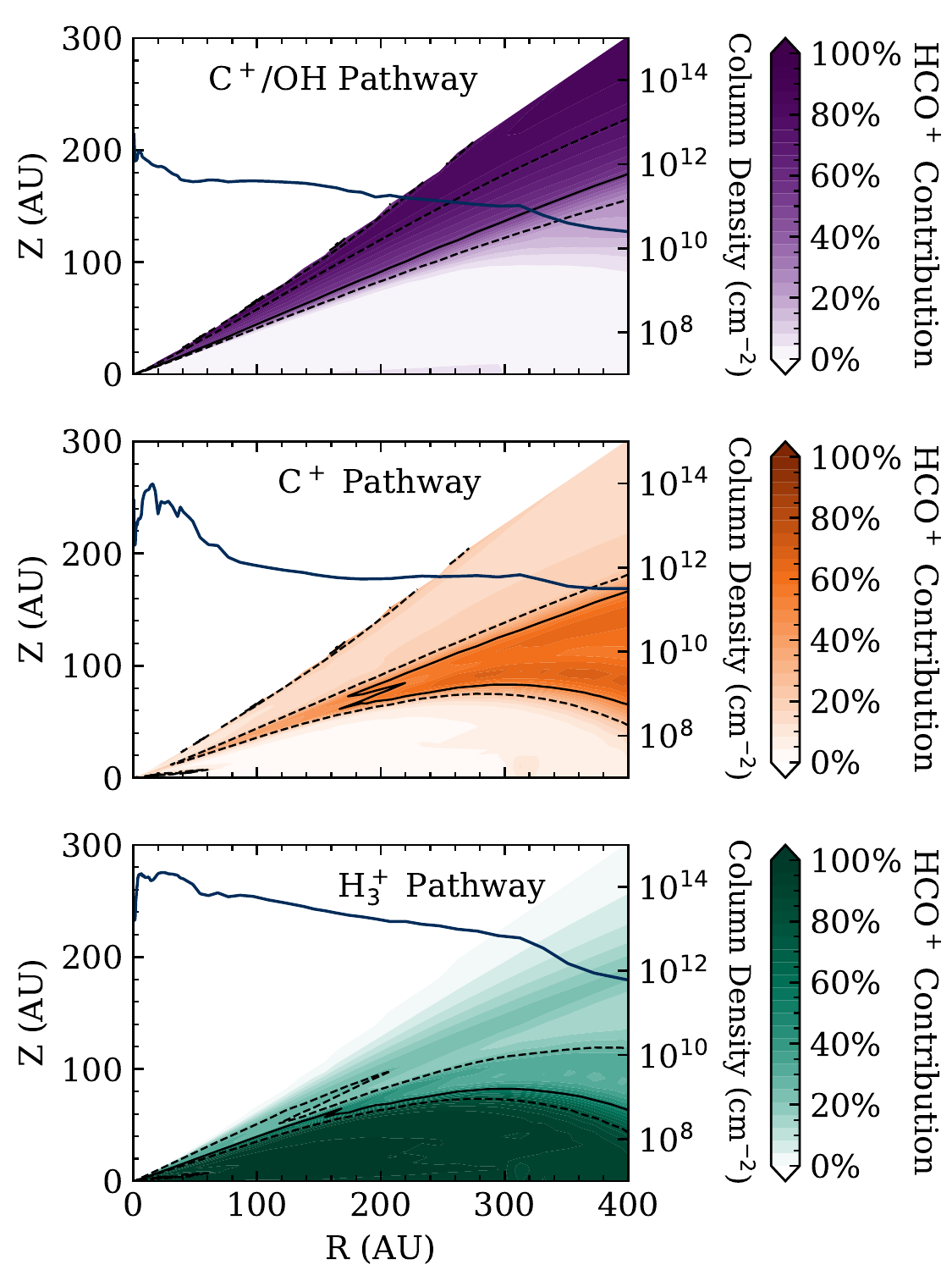}
\caption{Relative importance of the three dominant \hcop{} production pathways throughout the disk. Solid contours mark where each pathway accounts for 50\% of the total \hcop{} production rate at 0.5~Myr. Similarly, the dashed contours mark 25\% and 75\%. Additionally, dark blue curves indicate the column density attributed to each pathway.}
\label{fig:path_profiles}
\end{center}
\end{figure}

Our chemical models exhibit this expected layering. However, we see additional vertical structure --- especially in \hcop{} --- which we did not anticipate and ultimately attribute to enhanced UV ionization of disk material potentially caused by the large flaring angle of IM Lup. Figure \ref{fig:abundance} shows abundance profiles of CO, \hcop{}, and \nthp{} and highlights the effect of including radial depletion of CO. The canonical \hcop{}/\nthp{} layering is contained within $Z\le$ 50 au and $R\le$ 350 au. The unanticipated structures include (1) a band of \hcop{} and \nthp{} at $Z/R \sim$ 0.35 that arcs down to hit the midplane at  $R\sim$ 500 au, and (2) a diffuse \hcop{} band at $Z/R \sim$ 0.5.

\begin{figure*}[t]
   \begin{center}
    \includegraphics[width=.85\textwidth]{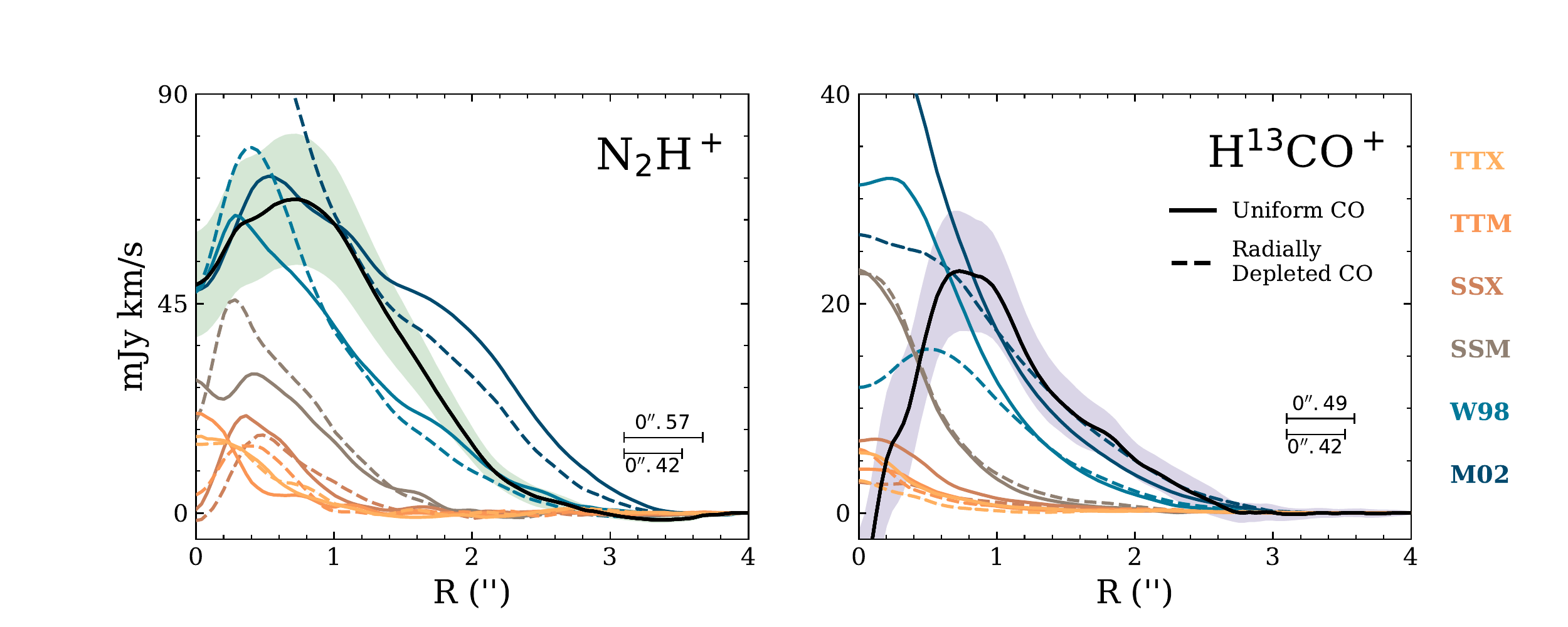}
    \caption{Results of single cosmic ray models. Solid black lines indicate observed \nthp{} 3--2 (left) and \hcisoop{} 3--2 (right) radial intensity profiles per beam, with shaded 1$\sigma$ uncertainty. Colored lines indicate model radial intensity profiles assuming a single cosmic ray template and uniform CO abundance (solid) or radially depleted initial CO abundance (dashed). We find that weak-CR (SSM, SSX, TTM, TTX) models unanimously produce under-bright emission, and while W98 and M02 models match well to \nthp{} emission, they produce excess \hcisoop{} 3--2 emission within $\sim1^{''}$.}
    \label{fig:singlecr}
    \end{center}
\end{figure*}

To explain the vertical structure, we analyzed the dominant reaction pathways forming \hcop{} and \nthp{} in our IM Lup model. For \hcop{} we find three mostly separable pathways operating in roughly distinct regions of the disk. In addition to the standard H$_3^+$ pathway, two strongly UV-driven pathways are operative in the diffuse upper/outer regions of the disk; one relying on UV-generated OH and C$^+$ above $Z/R \sim$ 0.4, and another relying on C$^+$ generated CH$_3^+$ between 0.3 $\lesssim$ $Z/R \lesssim$ 0.4.  Figure \ref{fig:path_diagram} shows a schematic of the three \hcop{} formation pathways, and Figure \ref{fig:path_profiles} shows how much each pathway contributes to the total \hcop{} production rate throughout the disk.

Contribution from each pathway was computed using rates of the dominant reactions forming \hcop{} and HOC$^+$ at each location in the disk, as described in Appendix~\ref{apdx:reac}. We find that the warm C$^+$ pathway (orange) dominates \hcop{} formation at the location of the $Z/R \sim$ 0.35 \hcop{} and \nthp{} bands, and the hot C$^+$ pathway involving OH (purple) dominates formation above $Z/R \sim$ 0.4, where the diffuse \hcop{} band exists. As for \nthp{}, we find that the dominant formation pathway throughout the disk is the canonical H$_3^+$ pathway, however we note that in the presence of N$_2$, HOC$^+$ can react with N$_2$ to form \nthp{}, and in the presence of OH, \nthp{} is destroyed. Together, these help to explain the presence of \nthp{} in the $Z/R \sim$ 0.35 band, and its absence above $Z/R \sim$ 0.4.

\vspace{1mm}
\subsection{Single-CR Models}\label{sec:singlecr}

Using the six CR models from Table \ref{tab:CRmods}, we created six disk models as discussed above. Observed and modelled radial intensity profiles for \nthp{} 3--2 and \hcisoop{} 3--2 are shown in Figure \ref{fig:singlecr}. Profiles are generated by azimuthally averaging 100 concentric elliptical annuli out to a 4$^{''}$ semi-major axis around the disk center.  For IM Lup, we adopt an inclination of 49$^\circ$ and a position angle of 144$^\circ$. Errors are computed empirically by applying the Keplerian mask to line-free channels many times and computing the standard deviation on a per-pixel basis to produce a moment-0 1$\sigma$ uncertainty map. For a given annulus, the error is estimated as the average uncertainty within the annulus divided by the the square root of the number of beams covering the annulus.

We find that only W98 and M02 models are able to produce sufficiently bright emission to be consistent with observations of IM Lup, especially in the outer regions of the disk. However, these models are overly bright in the inner region of the disk, leading us to explore models with reduced inner disk CR ionization. For these single-CR models, we computed $\chi^2$ goodness-of-fit and present results in Figure \ref{fig:chi} alongside those of hybrid models discussed in the next section.

\subsection{Multi-CR Models}\label{sec:multicr}

In order to match inner disk emission, we produced hybrid models utilizing two CR models (one for the inner disk, and one for the outer disk) transitioning sharply at a single radius. Merging of models occurred after the chemical modeling stage, and then merged model abundances were sent through the same synthetic observation pipeline discussed in Section \ref{sec:synth}. 

\begin{figure*}[t]
   \begin{center}
    \includegraphics[width=1.05\textwidth]{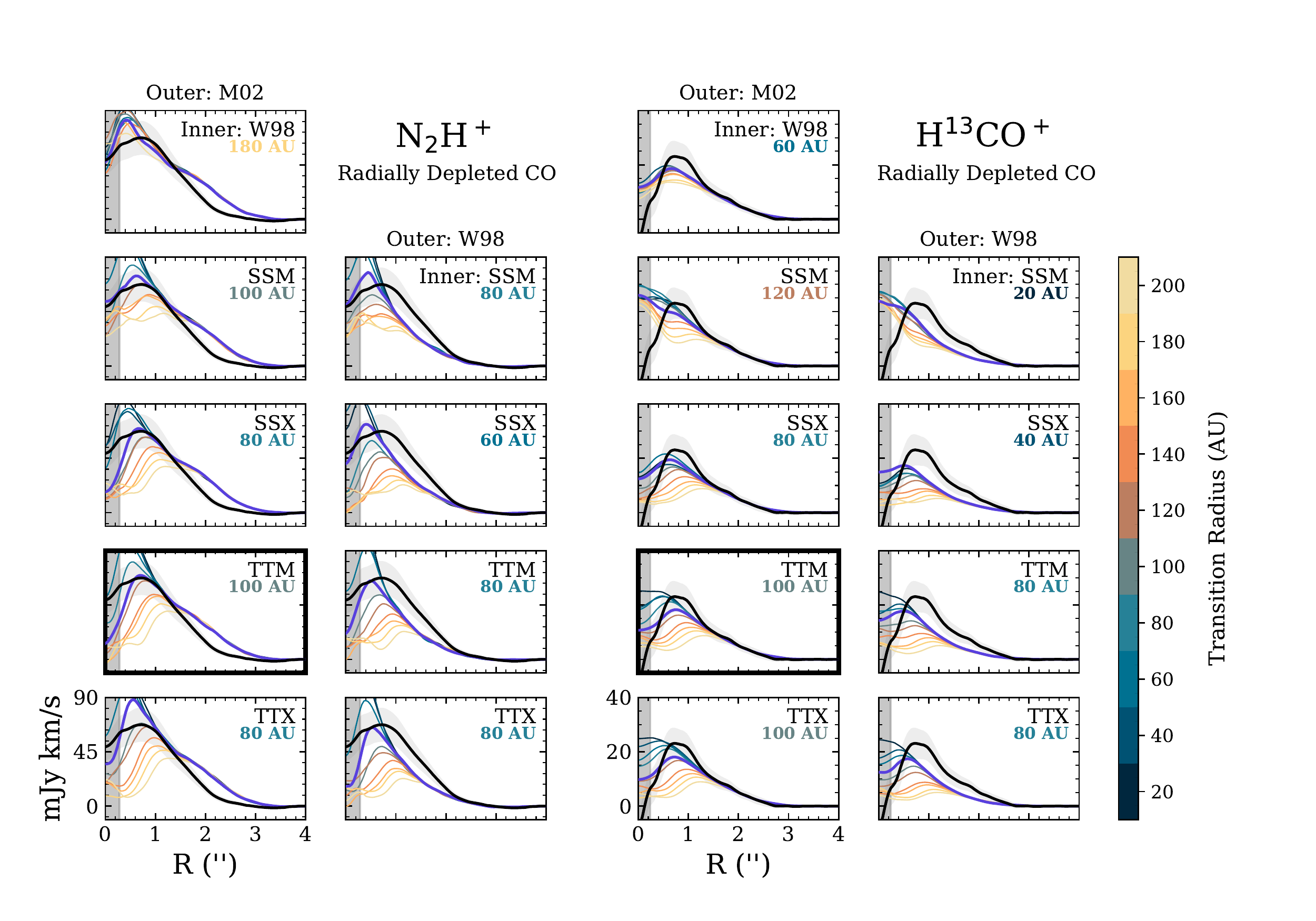}
    \vspace{-8mm}
    \caption{Radial intensity profiles for a subset of models with radially depleted initial CO. Each panel represents a different combination of inner and outer CR models. Line color indicates the assumed transition radius between low-CR inner disk and high-CR outer disk. For each panel, observed radial intensity is indicated in black, the best fit is highlighted in purple, and the best-fit transition radius is indicated in the upper right corner. See Appendix \ref{apdx:rprofs} for radial profiles of models with a uniform CO distribution.}
    \label{fig:radprf}
   \end{center}
\end{figure*}

We computed these hybrid models over a range of potential transition radii spanning from 20 au to 200 au, and only considered combinations of CR models with higher CR ionization in the outer disk than in the inner disk. A subset of the hybrid model profiles are shown in Figure~\ref{fig:radprf}.

\begin{figure*}[t]
    \begin{center}
    \includegraphics[width=1.05\textwidth]{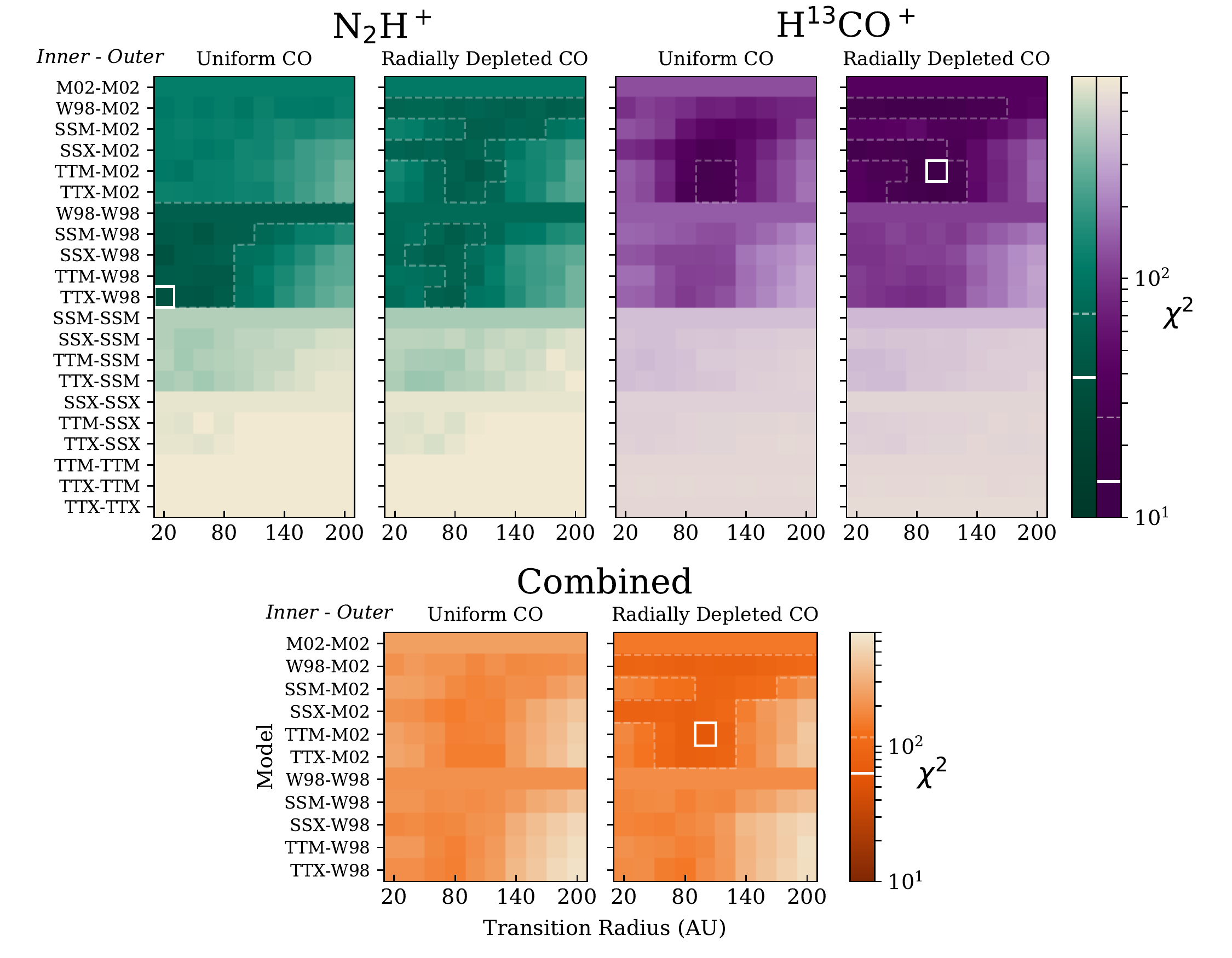}
    \vspace{-7mm}
    \caption{$\chi^2$ goodness-of-fit as a function of inner and outer CR model, transition radius, and initial CO distribution. The top row shows $\chi^2$ for individual lines; \nthp{} 3--2 (left two) and \hcisoop{} 3--2 (right two). The bottom row shows total $\chi^2$ for each model with outer M02 or W98. For each set of panels, the best-fit model is highlighted in white, and marginally good fits are indicated with dotted gray contours.\label{fig:chi}}
    \end{center}
\end{figure*}

To assess how well each model reproduces the observed emission, we compute $\chi^2$ between the observed and modelled radial intensity profiles, separately for the \nthp{} 3--2 and \hcisoop{} 3--2 lines. It is important to note that both lines are only detected above a 3$\sigma$ threshold out to a radius of $\sim$2$^{''}$, or 320 au, and therefore cannot be used to validate model emission beyond this point except to confirm that it is sufficiently faint. Profiles are computed out to a radius of 4$''$ (644 au), and we choose to ignore the inner 0$^{''}$\hspace{-1.5mm}.29 and 0$^{''}$\hspace{-1.5mm}.24, for \nthp{} 3--2 and \hcisoop{} 3--2 respectively, since the central beam may be affected by optically thick dust \citep{huang18}. These results are shown in Figure~\ref{fig:chi}. We achieve best-fit $\chi^2$ values of 14.1 and 38.5 for \hcisoop{} 3--2 and \nthp{} 3--2, respectively.

Individually, the best-fit models for both lines suggest that a model with a CR gradient is needed to reproduce the observed emission, with high CR ionization in the outer disk (W98 or M02), low CR ionization in the inner disk (TTX through SSM), and a transition somewhere between 20-120 au. However, the two molecular ions we probed favor different outer disk CR models; M02 is necessary to match outer disk \hcisoop{} 3--2 emission, but this same model produces slightly over-bright \nthp{} 3--2 beyond $\sim$ 1$^{''}$\hspace{-1.5mm}.5. Furthermore, \nthp{} 3--2 slightly favors our models with uniform CO distribution while \hcisoop{} 3--2 is fit much better by models with radially depleted CO. To determine the global best-fit model, we computed the total $\chi^2$ across both lines, and found that our model with TTM inside 100 au, M02 beyond 100 au, and a radially depleted initial CO distribution fits best to both \nthp{} and \hcisoop{} simultaneously, with a total $\chi^2$ value of 63.1. Individual and combined $\chi^2$ results are shown in figure \ref{fig:chi}.

The individual and combined results provide evidence for CR suppression in the inner disk, and the combined result suggests suppression within a radial scale of $\sim100$ au. We note that this result does not exclude the possibility that locally accelerated stellar cosmic rays (SCRs) are influencing disk chemistry. However it does set an upper limit on ionization, since the combination of galactic CRs and SCRs should not exceed an ionization rate of $\sim$10$^{-20}$~s$^{-1}$ from R $\sim$ 30--100 au near the midplane of this particular source. It is still possible that SCR ionization exceeds this rate and perhaps dominates ionization in the upper layers of the disk \citep{rab17,rodgers-lee17}, however this region has little impact on the observed ions, and therefore has a weak influence on the constraints we find.

\section{Discussion}\label{sec:disc}

By modelling the extended molecular ion emission in the IM Lup protoplanetary disk, we find that non-equilibrium chemical models with a single CR ionization rate do not reproduce observations well. Instead, we invoke a cosmic ray gradient to explain inner deficits in emission. In this section, we discuss how \hcisoop{} and \nthp{} emission morphologies support this interpretation (Section~\ref{sec:gaps}), the possible significance of the CR transition radius (Section~\ref{sec:whyswap}), and implications for ongoing migration of disk material (Section~\ref{sec:edens}).

\subsection{\nthp{} as a CO Snowline Tracer}\label{sec:gaps}

Contrary to previous interpretations of \nthp{} observations in disks, our results do not suggest \nthp{} is clearly tracing either CO or N$_2$ snow surfaces in IM Lup. We find that \nthp{} and \hcop{} remain abundant even in regions where N$_2$ and CO are predominantly frozen-out on grains, due to the high rate of CR desorption/ionization in the outer disk. Instead, we attribute the inner gap in \nthp{} and \hcisoop{} emission to a sharp drop in CR ionization as discussed above.

There is an extensive, robust history of modeling \nthp{} in disks \citep{vanthoff17,aikawa15} and using it to observationally infer the location of the CO snowline \citep{oberg15}, since it thrives in the absence of gas-phase CO. N$_2$ freezes out at temperatures $\sim$3--5 K cooler than CO, creating a sheath between the CO and N$_2$ snow-surfaces in which \nthp{} flourishes. At the midplane, this region spans from the CO snowline to the N$_2$ snowline, producing a ring of \nthp{} emission.

There are many instances where it has been shown that \nthp{} can indicate the location of the CO snowline. For example, \citet{qi19} identifies a number of disks with ring-like \nthp{} emission, and shows that for disks exhibiting a thin \nthp{} ring, the emission morphology is well-explained by CO/N$_2$ freeze-out with a thick Vertically Isothermal Region above the Midplane (VIRaM). 

However, IM Lup exhibits thick-ringed \nthp{} emission without a sharp outer edge. \citet{qi19} estimates an inner \nthp{} wall at 59 au, but does not fit an outer radius or model emission with their VIRaM model. Our chemical models have a CO snowline at $\sim$30 au, which is too small to produce the observed central deficit in \nthp{} emission and leads us to conclude that the inner \nthp{} deficit in IM Lup cannot be explained by freeze-out alone. 59 au, however, is consistent with lower estimates for the CR transition we find.

Since freeze-out alone cannot reproduce the \nthp{} morphology we observe, there must be an alternative explanation for the \nthp{} central gap, which we now attribute to a CR gradient. Furthermore, we see similar emission morphologies for both \hcisoop{} 3--2 and \nthp{} 3--2; both lines have peak emission at $\sim$1$''$ with a central deficit and emission out to $\sim$3$''$. This goes against the pure CO snowline driven chemistry interpretation, which should yield bright \hcisoop{} emission inside the \nthp{} deficit.

It remains unclear how many disks might exhibit this type of CR-induced \nthp{} ring versus the canonical snowline-induced ring, but we believe examining the morphologies of \textit{both} \nthp{} and \hcop{} is a reliable way to rule out the snowline explanation for individual disks. If morphologies of both species are similar, a CO snowline is a poor explanation for the location of peak \nthp{} emission.

\subsection{Interpretation of transition radius}\label{sec:whyswap}

A radial variation in the CR flux is perhaps less surprising when viewed in the context of our modern day Sun. Specifically, the Sun's heliopause approximately marks the boundary between the CR-modulated environment inside the domain of the solar wind and the interstellar CR environment beyond. As measured by {\em Voyager 1}, the heliopause radius is located at a distance of $\sim$120 au \citep[e.g.,][]{zhang15PhPl}. With IM Lup, perhaps we are also seeing the edge of influence of a stellar wind extending out to $\sim$100 au, beyond which there is a sharp increase in CR strength at an analogous T-Tauriopause.

The IM Lup disk is young and expansive. At an estimated age of 0.5 Myr, the gas disk extends out to $\sim$800 au, exposing much of the disk to the high-CR environment beyond the tentative 100 au T-Tauriopause. This location is interesting too because  it marks the edge of spiral structure identified by the DSHARP collaboration \citep{huang18spirals}, where build-up of material not efficiently transported by MRI could be leading to gravitational instability. 

That said, it may be the case that we are catching IM Lup at a convenient time in its evolution, when a large amount of disk material has yet to migrate within the potentially growing extent of the young star's wind. IM Lup is still shrouded in diffuse natal cloud material, which likely places restraints on the size of the T-Tauriopause, which is expected to become larger than 100 au given the mass-loss rate of the star. As the extent of IM Lup's wind grows and clears away cloud material, and as disk material continues to migrate inwards, it is possible that in the future the T-Tauriopause will fully encompass the disk, and the CR gradient we see today will vanish. Additional --- more detailed --- modeling of this system would be highly valuable to verify the radial ionization gradient we find, and investigate the plausibility that a T-Tauriopause could be contained within $\sim$100-150 au given the amount of natal cloud material that still remains.

Perhaps IM Lup only appears to be a special case because of how young the system is. Maybe all disks go through a period of CR gradient evolution early on while the disk is actively shrinking and winds are pushing natal cloud material outward. In comparison, \citet{cleeves15}'s ionization constraints on the $\sim$3--10 Myr-old TW~Hya protoplanetary disk did not find strong evidence for a CR gradient. However, given the lower SNR of their data, many lines were modeled with disk-integrated fluxes, and radial fits attempted for HCO$^+$ 3-2 and N$_2$H$^+$ 4-3 had relatively large errors. Thus only a single incident CR value was attempted in the fit. Nonetheless, it seems as if a single, lower CR rate (especially from 60--180 au) is a better match for TW Hya, suggesting that the region of exclusion may grow with time or fully encompass disks that are more mature. Additional high-resolution observations of disk ion emission covering a range of systems with varying ages and levels of remnant natal cloud material are critical to address this question, and we are eager to see if---and how many---other disks exhibit similar CR gradients.

In addition to better and more observations, future studies would also greatly benefit from more detailed modeling of the transition itself and a treatment of CR propagation across it. Our simplified sharp CR transition at a single radius serves primarily as a proof-of-concept, but in reality CRs enter the T-Tauriosphere from many lines-of-sight and could propagate some distance before being sufficiently deflected. For example, with our best-fit model we find that CRs traveling in along the midplane would reach the 100 au T-Tauriopause without significant attenuation. How these CRs would subsequently be deflected and attenuated depends on the magnetic field strength and configuration at the T-Tauriopause and would ultimately determine the width and shape of the transition from high ionization rates in the outer disk to low ionization rates in the inner disk.
\subsection{Mechanisms of Ongoing Migration}\label{sec:edens}

\begin{figure}[t]
   \begin{center}
    \includegraphics[width=0.48\textwidth]{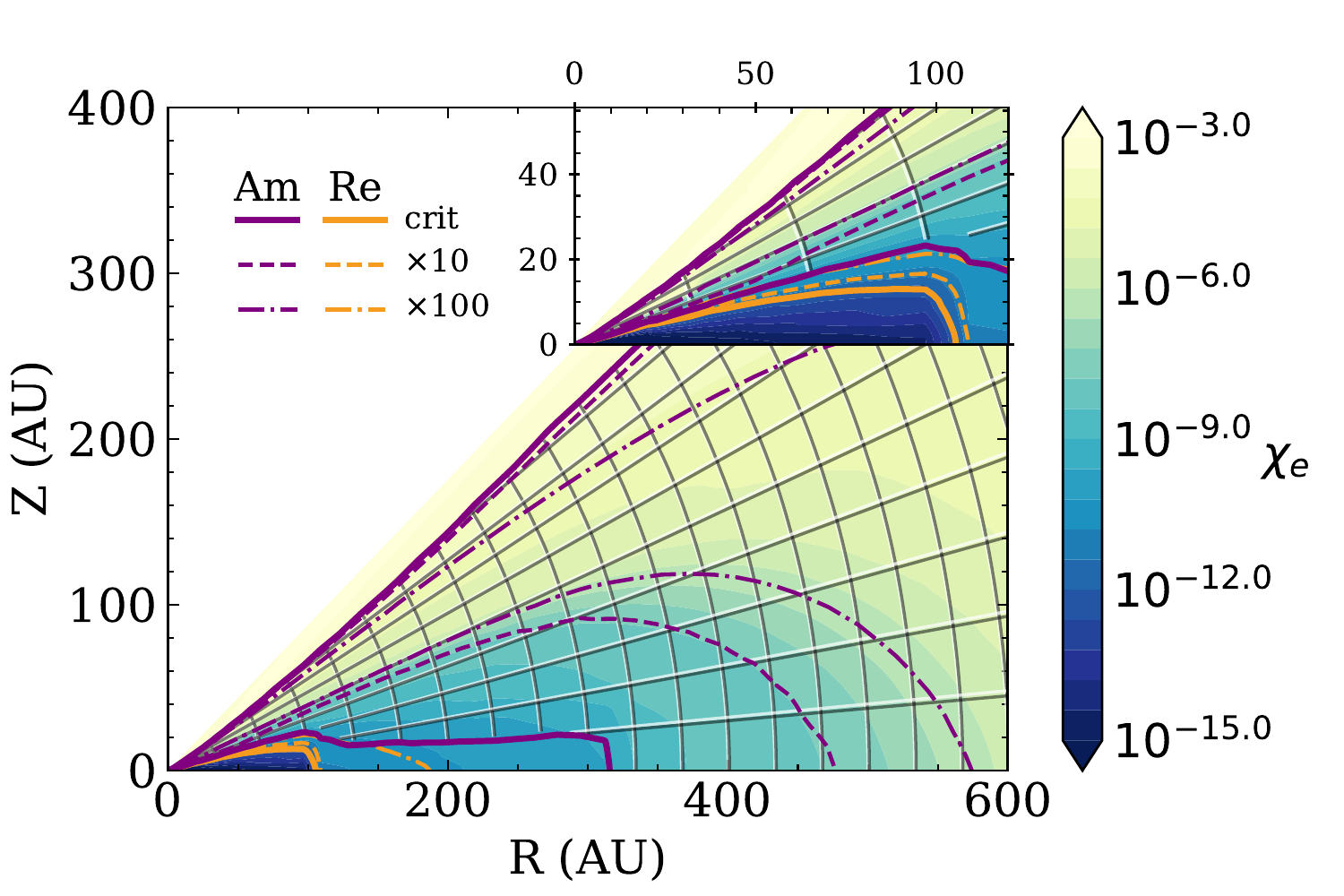}
    \caption{Charge fraction $\chi_e$ from our best-fit hybrid CR model. Contours indicate [1,10,100]$\times$ the critical values for Re (Re$_{crit}$ = 3000, orange) and Am (Am$_{crit}$ = 0.1, purple). Hatched regions indicate where both Re and Am criteria are satisfied, and it is feasible for the disk to be MRI active.\label{fig:edens}}
    \vspace{-2mm}
    \end{center}
\end{figure}

As part of our chemical modeling procedure, we can also retrieve the disk-wide electron abundance, $\chi_e$, which we show in Figure \ref{fig:edens}. With this, we can assess whether or not material is sufficiently coupled to stellar and disk magnetic fields for the Magneto-Rotational Instability (MRI) to facilitate the migration of disk material \citep[e.g.,][]{bai-stone11}. To identify regions of the disk that are feasibly MRI active, we evaluate two criterion; the magnetic Reynolds number,
\begin{equation}
    {\rm Re} \equiv \frac{c_sh}{D} \approx 1 \left(\frac{\chi_e}{10^{-13}}\right) \left(\frac{T}{100 {\rm K}}\right)^{1/2} \left(\frac{a}{{\rm au}}\right)^{3/2},\\
\end{equation}
which parameterizes how well the charged disk is coupled to magnetic fields, and also the ambipolar diffusion coefficient, 
\begin{equation}
    {\rm Am} \equiv \frac{n_{i}\beta_{in}}{\Omega} \approx 1 \left(\frac{\chi_{i}}{10^{-8}}\right) \left(\frac{n_{tot}}{10^{10} {\rm cm}^{-3}}\right) \left(\frac{a}{{\rm au}}\right)^{3/2},
\end{equation}
which parameterizes how well the neutral disk is coupled to disk ions \citep{perez-becker11b}. Here $c_s$ is the sound speed, $h$ is the disk scale height, $D$ is the magnetic diffusivity, $\beta_{in}$ is the collisional rate coefficient between charged and neutral particles, and $\Omega$ is the orbital frequency.

Following \citet{cleeves13a}, we require Re~$>$~3000 and Am~$>$~0.1 in order for a region of the disk to be deemed MRI active. These criteria come from models run by \cite{flock12} indicating Re must exceed a value of 3000 to sustain turbulence, and from \cite{bai-stone11} who find that in weakly magnetic disks, the coefficient Am must exceed 0.1 to maintain sufficiently frequent ion-neutral collisions.

Figure \ref{fig:edens} shows the electron fraction $\chi_e$ for our best-fit hybrid CR disk model, with hatching to indicate regions where the model is feasibly MRI active. We find that the entire disk is MRI active beyond the edge of the dust disk at 313 au, and interior to this,  MRI is active above $\sim$20 au. However, the disk midplane is predominantly MRI inactive. The lack of efficient magnetically-assisted transport in this region may be responsible for the low mass accretion rate inferred for this disk \citep[10$^{-11}$ M$_\odot$ yr$^{-1}$,][however, see also \citet{acala2017} as the rate is perhaps higher or quite variable]{guenther10,siwak16}, and if it is leading to a buildup of disk material, could help explain the existence of spiral structure within 100 au \citep{huang18spirals}.

\section{Conclusions}\label{sec:conclusions}

We present observations of N$_2$H$^+$ 3--2 and H$^{13}$CO$^+$ 3--2 emission toward the IM Lup protoplanetary disk with the goal of fitting these data using detailed chemical models and observational post processing. Our main conclusions can be summarized as follows:

\begin{itemize}
    \item The outer disk of IM Lup exhibits high CR ionization rates comparable to the diffuse ISM. In contrast, the inner disk exhibits low CR ionization rates, which is necessary to explain inner deficits in both \nthp{} 3--2 and \hcisoop{} 3--2 emission (which are too wide to be fully explained by opaque dust).  The transition between high and low CR environments occurs near a radial scale $R\sim$ 80 -- 100~au. Midplane ionization rates are approximately $\zeta_{CR}$ $\lesssim$ 10$^{-20}$ s$^{-1}$ out to 100 au, and $\zeta_{CR}$ $\gtrsim$ 10$^{-17}$ s$^{-1}$ between 100 au and 300 au (beyond which we do not detect molecular ion emission). This transition radius is co-spatial with the edge of spiral structure seen in DSHARP observations \citep{huang18spirals}. The steep increase in incident CR rate could signify the edge of a T Tauriosphere, i.e., a stellar wind induced boundary analogous to the heliosphere in our solar system. 
    
    \item It is feasible that IM Lup is MRI active at all locations of the disk that are either 1) above $Z/R \sim$0.25, or 2) more than $\sim$20 au above the midplane, or 3) beyond the mm dust disk edge at a radius of 313 au ($\sim2''$). The largely MRI inactive midplane may explain IM Lup's low mass accretion rate. If this property is leading to the build-up of disk material within the CR transition radius, it could also explain the emergence of spiral structure within 100 au reported by \citet{huang18spirals}.
    
    \item The disk in IM Lup plays host to an enhanced UV-driven ionization chemistry, likely due to its size and high flaring angle. We identify two enhanced reaction pathways dominating the formation of \hcop{}, which give rise to \hcop{} and \nthp{} vertical structure above $Z/R \sim$ 0.3 and affect their midplane abundances beyond $R\sim$ 450 au in our models.
\end{itemize}

These results are interesting in relation to previous chemically constrained ionization in the TW Hya protoplanetary disk. \citet{cleeves15} reported an upper limit on the disk-averaged CR ionization rate of $\zeta_{CR}\lesssim 10^{-19}$ s$^{-1}$ per H$_2$. Due to limited spatial resolution, the TW Hya CR models were primarily constrained with disk integrated line fluxes; only \nthp{} 4--3 and \hcop 3--2 provided some spatial constraints. However, the TW Hya disk's \nthp{} only constrained ionization within the inner 120 au, and therefore it was not possible to accurately resolve ionization constraints radially with these data. As a result, based on the analysis presented here, it is not surprising that TW Hya has an intermediate value for its disk-averaged ionization level that falls between IM Lup's low inner disk and high outer disk. This finding highlights the need for sensitive, spatially resolved multi-molecule maps to make detailed constraints on ionization -- an essential parameter underpinning disk physics and chemistry.

\acknowledgments We are grateful to the referee whose thorough comments improved the manuscript. We are also grateful to E. A. Bergin and R. Visser in the preparation of the observing proposals through which these data were obtained. This paper makes use of the following ALMA data: ADS/JAO.ALMA\#2013.1.00694.S and ADS/JAO.ALMA\#2013.1.00226.S. ALMA is a partnership of ESO (representing its member states), NSF (USA) and NINS (Japan), together with NRC (Canada), MOST and ASIAA (Taiwan), and KASI (Republic of Korea), in cooperation with the Republic of Chile. The Joint ALMA Observatory is operated by ESO, AUI/NRAO and NAOJ. The National Radio Astronomy Observatory is a facility of the National Science Foundation operated under cooperative agreement by Associated Universities, Inc. The modeling conducted in this paper was carried out on the University of Virginia's Rivanna High Performance Computing Cluster, for which we are grateful to have access to. LIC gratefully acknowledges support from NASA Astrophysics Theory Program 80NSSC20K0529, the David and Lucille Packard Foundation, and Johnson \& Johnson's WiSTEM2D Award, which supported this work. FCA is supported in part by the NASA Exoplanets Research Program (grant number NNX16AB47G). ZYL is supported in part by NSF AST-1910106 and NASA 80NSSC20K0533. We also acknowledgement the community of support and collaboration from the Virginia Initiative on Cosmic Origins.

\appendix

\section{HCO$^+$ Formation Route Calculation}\label{apdx:reac}

In Section~\ref{sec:chempath}, we estimate the approximate fractional contribution to \hcop{} formation from various pathways at 0.5 Myr (see Figure \ref{fig:path_profiles}). Since many pathways loop back on themselves or are interconnected, we attempt to break the key reactions into three broader channels (as shown in Figure \ref{fig:path_diagram}) to characterize the root source of ionization resulting in \hcop{} production throughout the disk. The three pathways we identify are \textbf{A)} a UV-driven, hot C$^+$/OH pathway (purple), \textbf{B)} a UV-driven, warm C$^+$ pathway (orange), and \textbf{C)} a CR/X-ray-driven H$_3^+$ pathway (green). Additionally, all three pathways have offshoots through the intermediate product HOC$^+$, which is efficiently converted into the more energetically favorable \hcop{} through a reaction we denote as D.

To compute pathway aggregate rates, we sum the rates of endpoint reactions resulting in \hcop{} for each pathway, and add a fraction of the HOC$^+$ conversion rate (reaction D) proportional to that pathway's contribution to HOC$^+$ production, as shown below:

\begin{alignat*}{4}
    A1&:& \textrm{CO$^+$} + \textrm{H$_2$} &\rightarrow \textrm{H} + \textbf{HCO$^+$} \\
    A1^\prime&:& \textrm{CO$^+$} + \textrm{H$_2$} &\rightarrow \textrm{H} + \textbf{HOC$^+$} \\ \\
    B1&:& \textrm{C$^+$} + \textrm{H$_2$O} &\rightarrow \textrm{H} + \textbf{HCO$^+$} \\
    B1^\prime&:& \textrm{C$^+$} + \textrm{H$_2$O} &\rightarrow \textrm{H} + \textbf{HOC$^+$} \\
    B2&:& \textrm{CH$_3^+$} + \textrm{O} &\rightarrow \textrm{H$_2$} + \textbf{HCO$^+$} \\
    B2^\prime&:& \textrm{CH$_3^+$} + \textrm{O} &\rightarrow \textrm{H$_2$} + \textbf{HOC$^+$} \\
    B3&:& \textrm{CH$_5^+$} + \textrm{CO} &\rightarrow \textrm{CH$_4$} + \textbf{HCO$^+$} \\ \\
    C1&:& \textrm{H$_3^+$} + \textrm{CO} &\rightarrow \textrm{H$_2$} + \textbf{HCO$^+$} \\
    C1^\prime&:& \textrm{H$_3^+$} + \textrm{CO} &\rightarrow \textrm{H$_2$} + \textbf{HOC$^+$} \\
    C2&:& \textrm{N$_2$H$^+$} + \textrm{CO} &\rightarrow \textrm{N$_2$} + \textbf{HCO$^+$} \\ \\
    D&:& \hspace{1mm}\textbf{HOC$^+$} + \textrm{H$_2$} &\rightarrow \textrm{H$_2$} + \textbf{HCO$^+$}
\end{alignat*}
\begin{align}
    r_{\textrm{HOC$^+$}} &\equiv r_{\textrm{A1}^\prime}+r_{\textrm{B1}^\prime}+r_{\textrm{B2}^\prime}+r_{\textrm{C1}^\prime} \nonumber\\ \nonumber\\
    r(\textrm{C$^+$/OH Pathway}) &= r_{\textrm{A1}} +  r_{\textrm{D}}\frac{r_{\textrm{A1}^\prime}}{r_{\textrm{HOC$^+$}}} \\
    r(\textrm{C$^+$ Pathway}) &= r_{\textrm{B1}} + r_{\textrm{B2}} + r_{\textrm{B3}} + r_{\textrm{D}}\frac{r_{\textrm{B1}^\prime} + r_{\textrm{B2}^\prime}}{r_{\textrm{HOC$^+$}}} \\
    r(\textrm{H$_3^+$ Pathway}) &= r_{\textrm{C1}} + r_{\textrm{C2}} + r_{\textrm{D}}\frac{r_{\textrm{C1}^\prime}}{r_{\textrm{HOC$^+$}}} \\
    \textrm{$^*r$ denotes a }& \textrm{rate in cm$^{-3}$ s$^{-1}$.} \nonumber
\end{align}

Finally, the fractional contribution of each pathway is computed by dividing the rate of that pathway by the summed rate of all three pathways. This is what is shown in Figure \ref{fig:path_profiles}.

\newpage
\section{Complementary Multi-CR Radial Profiles}\label{apdx:rprofs}

In Section \ref{sec:multicr}, intensity profiles were only shown for models with radially depleted initial CO distributions, favored by our global best-fit model. Here, for completeness, we show complementary radial profiles for models with a uniform initial CO distribution.

\begin{figure*}[h]
   \begin{center}
    \includegraphics[width=1.05\textwidth]{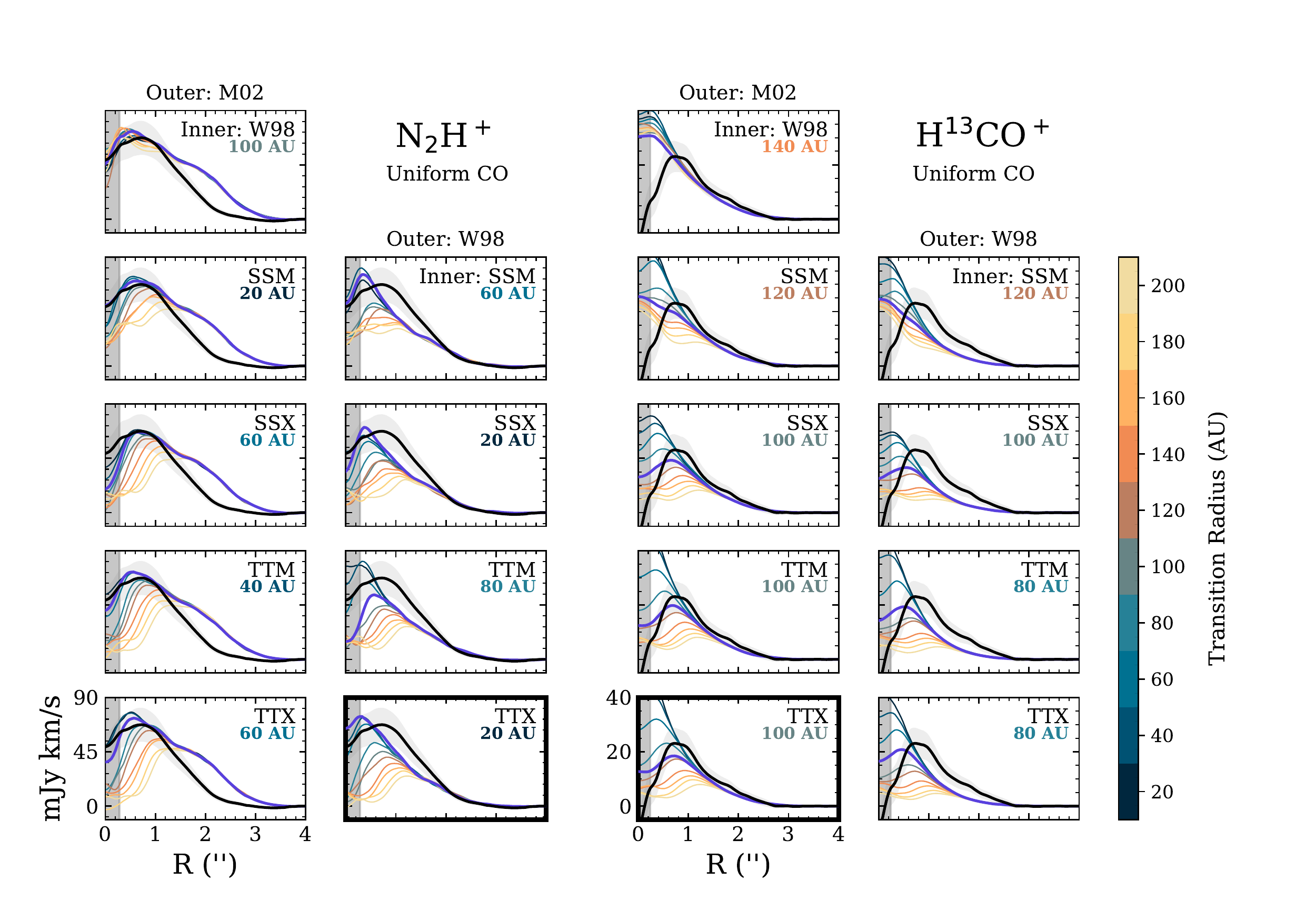}
    \vspace{-8mm}
    \caption{Same as Figure \ref{fig:radprf}, but for models with a uniform initial CO distribution.}
    \label{fig:radprf_ap}
   \end{center}
\end{figure*}

\newpage
\section{Model Column Densities}\label{apdx:rprofs}

Shown below are vertical column densities for the chemical models we used to produce hybrid CR gradient models. The sharp transition seen in some models at $\sim$30~au represents the CO snowline in our models, interior to which the midplane \nthp{} abundance plummets and initial CO abundance returns to ISM levels in radially depleted CO models.

Hybrid models discussed in section \ref{sec:multicr} are created using two of the single-CR models shown below, and transitioning from the inner CR model to the outer CR model sharply at a single radius. Thus, the column density of hybrid models also sharply transitions from the inner CR model to the outer CR model at the transition radius. To illustrate this, we show the vertical column density of our global best-fit model, with TTM CRs in the inner disk and M02 CRs in the outer disk, with a transition at 100 au.

\begin{figure*}[h]
   \begin{center}
    \includegraphics[width=1.05\textwidth]{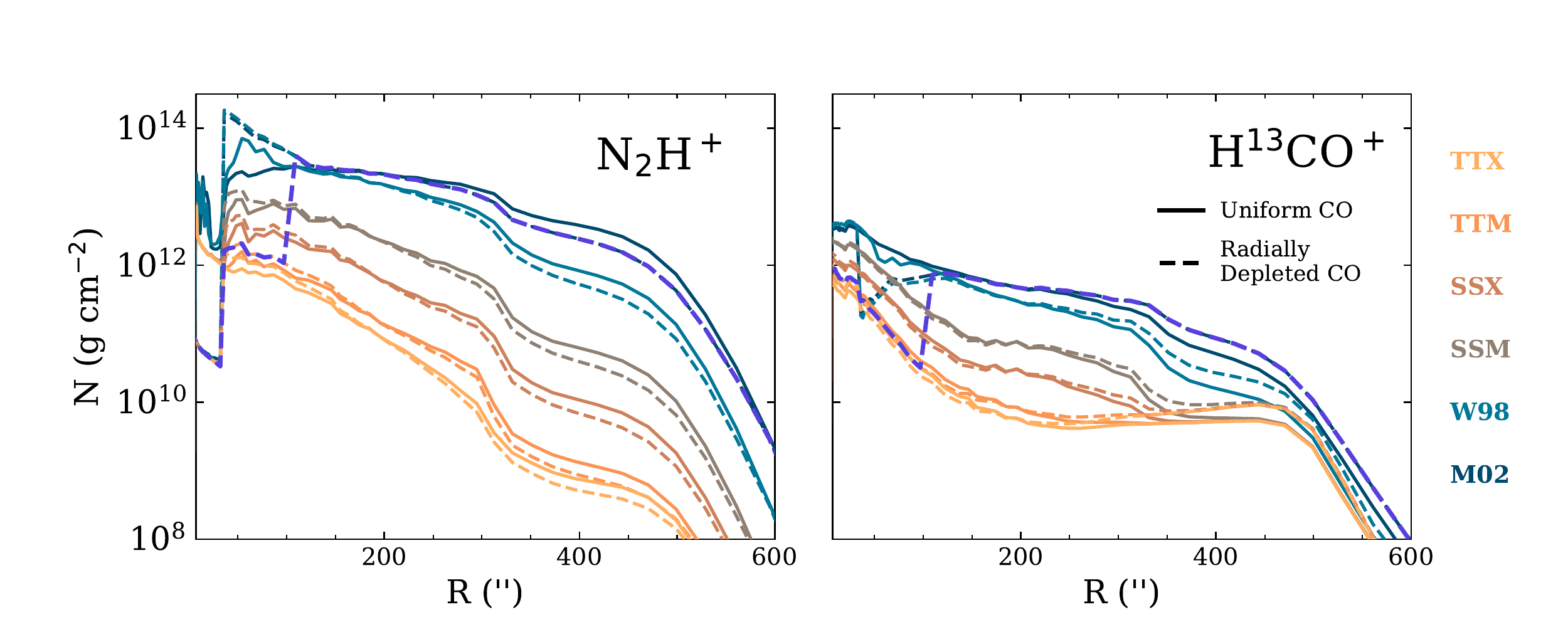}
    \vspace{-8mm}
    \caption{Vertical column density as a function of radius for the single-CR models used in our analysis, both with uniform (solid) and radially depleted (dotted) initial CO abundance. Indicated in purple is the column density of the best-fit hybrid CR model, which uses the TTM CR model within 100 au and the M02 model beyond 100 au.}
    \label{fig:radprf_ap}
   \end{center}
\end{figure*}

\section{Results with 1 Myr Model}\label{apdx:rprofs}

The age of IM Lup remains relatively uncertain, with upper estimates reaching as old as $\sim$2 Myr. To verify that our results are robust against the age we adopt for IM Lup, we carried out our analysis with a system age of 1 Myr and show results below.

We chose not to test older system ages (e.g. 1.5 or 2 Myr) because the physical disk model we use \citep{cleeves16} is fine-tuned with an age of 0.5 Myr. Since our chemical modeling is performed without dynamic physical disk evolution, models we compute far beyond 0.5 Myr cannot account for additional physical evolution of the disk, and give rise to unrealistic chemical artifacts, like excess sequestration of gas-phase CO into hydrocarbon ices that form on grains whose subsequent migration we do not model.

Qualitatively, we recover the same result with an age of 1 Myr as we did with an age of 0.5 Myr (see Figures \ref{fig:radprf_ap} and \ref{fig:chi_ap}); the closest-fit models have high CR ionization in the outer disk, low CR ionization in the inner disk, and a transition around 60-120 au. The best-fit model now favors a slightly weaker inner disk CR ionization (TTX), but we come to the same general conclusions with either assumed age.

\begin{figure*}[h]
   \begin{center}
    \includegraphics[width=0.9\textwidth]{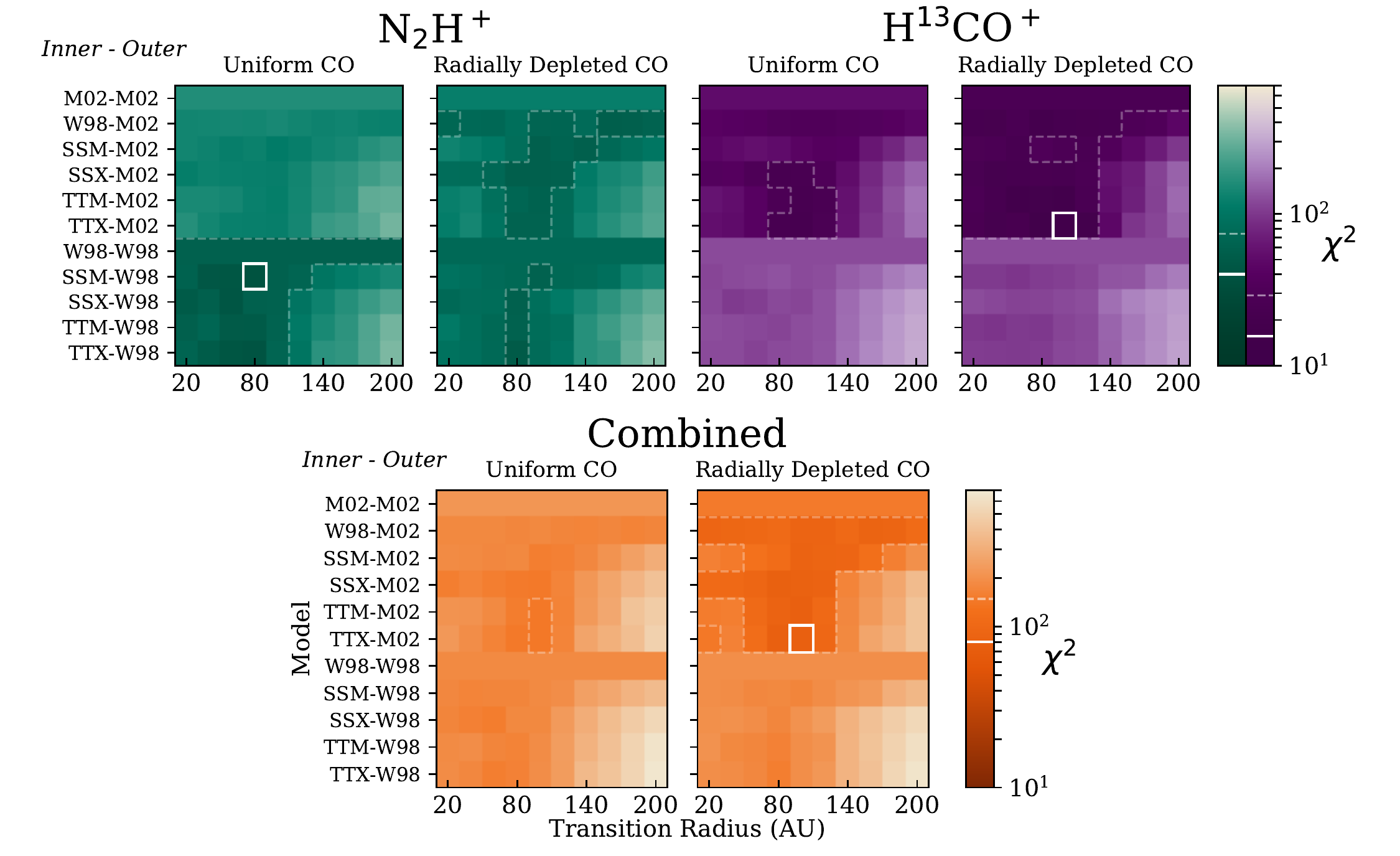}
    \vspace{-4mm}
    \caption{Same as Figure \ref{fig:chi}, but with models computed to an age of 1 Myr, instead of 0.5 Myr. To save computational resources, we only conducted this analysis at 1 Myr for models with W98 or M02 outer CR ionization.}
    \label{fig:radprf_ap}
   \end{center}
\end{figure*}

\begin{figure*}[h]
   \begin{center}
    \includegraphics[width=0.9\textwidth]{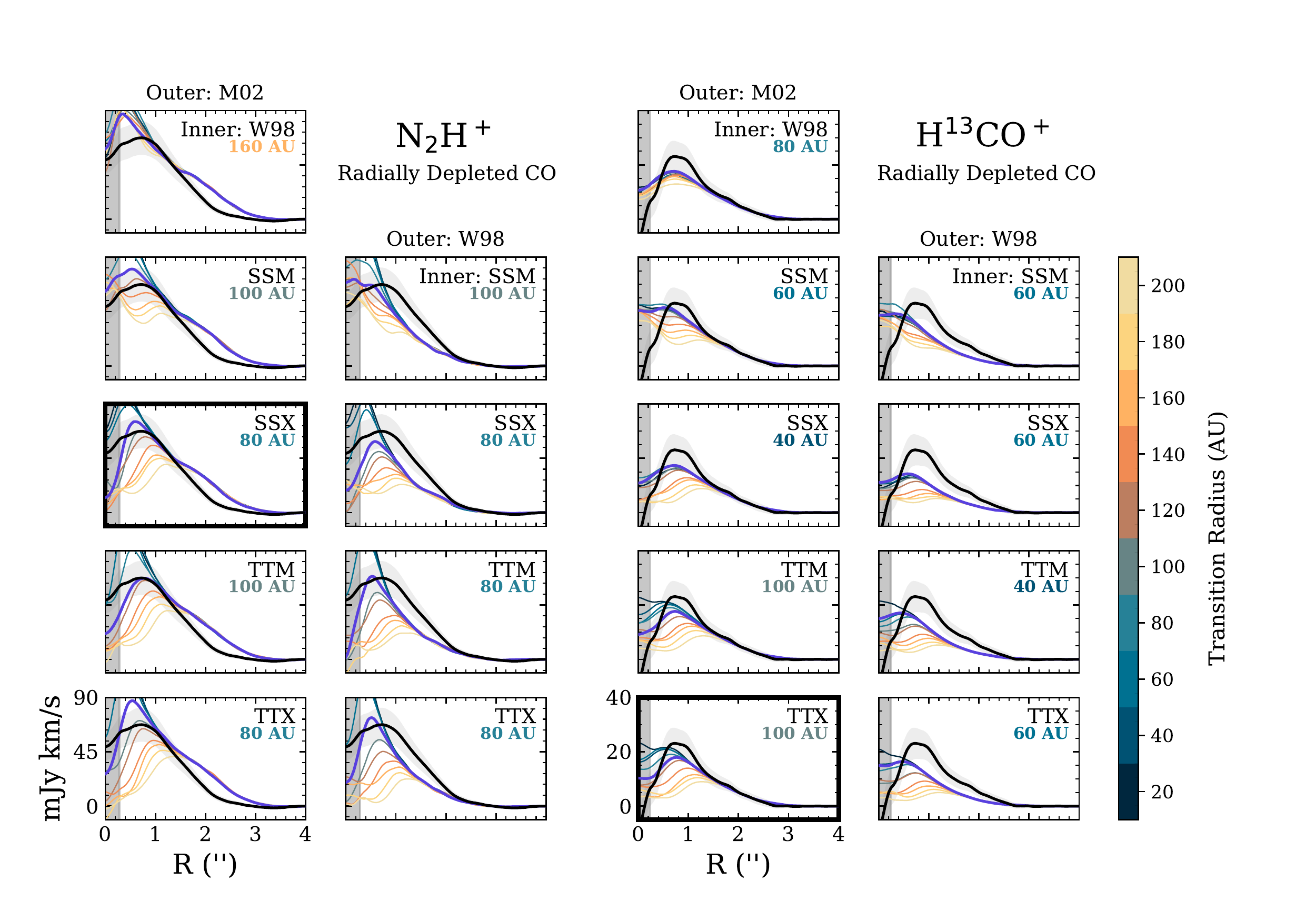}
    \vspace{-8mm}
    \caption{Same as Figure \ref{fig:radprf}, but with models computed to an age of 1 Myr, instead of 0.5 Myr.}
    \label{fig:chi_ap}
   \end{center}
\end{figure*}

\end{document}